\documentclass[conference]{IEEEtran}
\makeatletter
\def\ps@headings{%
\def\@oddhead{\mbox{}\scriptsize\rightmark \hfil \thepage}%
\def\@evenhead{\scriptsize\thepage \hfil \leftmark\mbox{}}%
\def\@oddfoot{}%
\def\@evenfoot{}}
\makeatother 
\usepackage{cite}
\usepackage{graphicx,amssymb,lineno}
\usepackage{cite}
\usepackage{amsmath,amsfonts,amssymb}
\usepackage[usenames]{color}
\usepackage{amsmath}
\usepackage[ruled,vlined]{algorithm2e}
\usepackage{subfigure}
\usepackage{multirow}
\usepackage{array}

\begin{document}
\title{Virtual Machine Migration Planning in Software-Defined Networks
}

\author{\IEEEauthorblockN{Huandong~Wang\IEEEauthorrefmark{1},Yong~Li\IEEEauthorrefmark{1},
Ying~Zhang\IEEEauthorrefmark{2},
 Depeng~Jin\IEEEauthorrefmark{1}
 }
\IEEEauthorblockA{\IEEEauthorrefmark{1}Tsinghua National Laboratory for Information Science and Technology\\
Department of Electronic Engineering, Tsinghua University, Beijing 100084, China\\}
\IEEEauthorblockA{\IEEEauthorrefmark{2}Ericsson Research 
}
Email: liyong07@tsinghua.edu.cn} %

\maketitle

\begin{abstract}
Live migration is a key technique for virtual machine (VM) management in data center networks,
which enables flexibility in resource optimization, fault tolerance, and load balancing.
Despite its usefulness,
the live migration still introduces performance degradations during the migration process.
Thus, there has been continuous efforts in reducing the migration time
in order to minimize the impact. From the network's perspective,
the migration time is determined by the amount of data to be migrated
and the available bandwidth used for such transfer.
In this paper, we examine the problem of how to schedule the migrations
and how to allocate network resources for migration
when multiple VMs need to be migrated at the same time.
We consider the problem in the Software-defined Network (SDN) context
since it provides flexible control on routing.

More specifically, we propose a method that computes the optimal migration sequence
and network bandwidth used for each migration.
We formulate this problem as a mixed integer programming, which is NP-hard.
To make it computationally feasible for large scale data centers,
we propose an approximation scheme via linear approximation
plus fully polynomial time approximation,
and obtain its theoretical performance bound.
Through extensive simulations, we demonstrate that our fully polynomial time approximation (FPTA) algorithm has a good performance compared with the optimal solution and two state-of-the-art algorithms.
That is,
our proposed FPTA algorithm approaches to the optimal solution with less than 10\% variation
and much less computation time.
Meanwhile,
it reduces the total migration time and the service downtime
by up to 40\% and 20\% compared with the state-of-the-art algorithms, respectively.
\end{abstract}

\section{Introduction}\label{sec:Introduction}
The modern cloud computing platform has leveraged virtualization to achieve economical multiplexing benefit while achieving isolation and flexibility simultaneously.
Separating the software from the underlying hardware, virtual machines (VMs) are used to host various cloud services \cite{30}.
VMs can share a common physical host as well as be migrated from one host to another. Live migration, $i.e.$, moving VMs from one physical machine to another without disrupting services, is the fundamental technique that enables flexible resource management in the virtualized data centers.
By adjusting the locations of VMs dynamically, we can optimize various objective functions to provide better services, such as improving performance, minimizing failure impact and reducing energy consumption \cite{1}.

While there are continuous efforts on the optimal VM placements to reduce network traffic\cite{28,29}, VM migration has received relatively less attention. We argue that careful planning of VM migration is needed to improve the system performance. Specifically, the migration process consumes not only CPU and memory resources at the source and the migrated target's physical machines \cite{29,31}, but also the network bandwidth on the path from the source to the destination\cite{29}. The amount of available network resource has a big impact on the total migration time, $e.g.$,
it takes longer time to transfer the same size of VM image with less bandwidth. As a consequence, the prolonged migration time should influence the application performance. Moreover, when multiple VM migrations occur at the same time, we need an intelligent scheduler to determine which migration tasks to occur first or which ones can be done simultaneously, in order to minimize the total migration time.

More specifically, there can be complex interactions between different migration tasks. While some independent migrations can be performed in parallel, other migrations may share the same bottleneck link in their paths. In this case, performing them simultaneously leads to longer total migration time.
In a big data center, hundreds of migration requests can take place in a few minutes \cite{38},
where the effect of the migration order becomes more significant.
Therefore, we aim to design a migration plan to minimize the total migration time
by determining the orders of multiple migration tasks, the paths taken by each task, and the transmission rate of each task.

There have been a number of works on VM migration in the literature. Work \cite{12,13} focused on minimizing migration cost by determining an optimal sequence of migration. However, their algorithms were designed under the model of one-by-one migration, and thus cannot perform migration in parallel simultaneously, leading to a bad performance in terms of the total migration time.
Bari $et\ al.$ \cite{11} also proposed a migration plan of optimizing the total migration time by determining the migration order. However, they assumed that the migration traffic of one VM only can be routed along one path in their plan. Compared with single-path routing, multipath routing is more flexible and can provide more residual bandwidth.
Thus, we allow multiple VMs to be migrated simultaneously via multiple routing paths in our migration plan.

In this paper, we investigate the problem of how to reduce the total migration time in Software Defined Network (SDN) scenarios\cite{10,13}. We focus on SDN because with a centralized controller, it is easier to obtain the global view of the network, such as the topology, bandwidth utilization on each path, and other performance statistics. On the other hand, SDN provides a flexible way to install forwarding rules so that we can provide multipath forwarding between the migration source and destination. In SDN, the forwarding rules can be installed dynamically and we can split the traffic on any path arbitrarily.
We allow multiple VMs to be migrated simultaneously via multiple routing paths.
The objective of this paper is to develop a scheme that is able to optimize the total migration time
by determining their migration orders and transmission rates.
Our contribution is threefold, and is summarized as follows:
\begin{itemize}
\item We formulate the problem of VM migration from the network's perspective, which aims to reduce the total migration time by maximizing effective transmission rate in the network,
    which is much easier to solve than directly minimizing the total migration time.
    Specifically, we formulate it as a mixed integer programming (MIP) problem, which is NP-hard.
\item We propose an approximation scheme via
linear approximation plus fully polynomial time approximation, termed as FPTA algorithm, to solve the formulated problem in a scalable way.
Moreover, we obtain its theoretical performance bound.
\item By extensive simulations,
we demonstrate that our proposed FPTA algorithm
achieves good performance in terms of reducing total migration time,
which reduces the total migration time by up to 40\% and shorten the service downtime by up to 20\% compared with the state-of-the-art algorithms.
\end{itemize}

The rest of the paper is organized as follows. In
Section \ref{sec:system}, we give a high-level overview of our system,
and formulate the problem of maximizing effective transmission rate in the network.
In Section \ref{sec:Approximation}, we propose an approximation scheme composed of
a linear approximation and a fully polynomial time approximation to solve the problem.
Further, we provide its performance bound.
In Section \ref{sec:Evaluation}, we evaluate the performance of our solution
through extensive simulations. After presenting related works
in Section \ref{sec:RelatedWorks}, we draw our conclusion in Section \ref{sec:Conclusion}.

\section{System Model and Problem Formulation}\label{sec:system}

\subsection{System Overview}\label{sec:S1}

\begin{figure} [t]
\begin{center}
\includegraphics*[width=8cm]{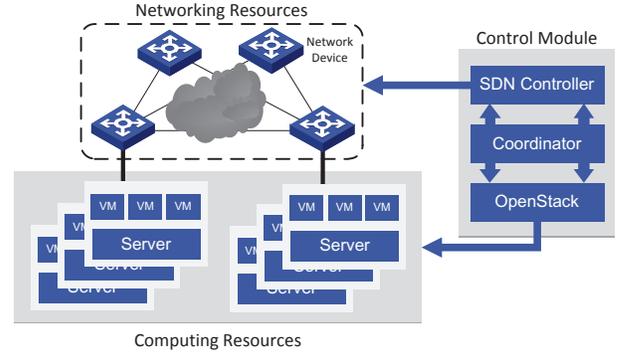}
\end{center}
\caption{System Overview} \label{fig:fig1}
\end{figure}

We first provide a high-level system overview in this section. As shown in Fig.~\ref{fig:fig1},  in such a network, all networking resources are under the control of the SDN controller, while all computing resources are under the control of some cloud management system, such as OpenStack.
Our VM migration plan runs at the Coordinator and it is carried out
via the OpenStack and SDN controller.

More specifically, devices in the network, switches or routers, implement forwarding according to their obtained forwarding tables and do some traffic measurement.
The SDN controller uses a standardized protocol, OpenFlow, to communicate with these network devices, and gather link-state information measured by them.
Meanwhile, the SDN controller is responsible for computing the forwarding tables for all devices.
On the other hand, the cloud controller, OpenStack, is responsible for managing all computing and storage resources.
It keeps all the necessary information about virtual machines and physical hosts,
such as the memory size of the virtual machine, the residual CPU resource of the physical host.
Meanwhile, all computing nodes periodically report their up-to-date information to it.
Besides, OpenStack also provides general resource management functions such as placing virtual machines, allocating storage, etc.

The processes of VM migration are described as follows.
Firstly, migration requests of applications are sent to the Coordinator. Based on the data collected from the OpenStack and SDN controller, the VM migration plan outputs a sequence of the VMs to be migrated with their corresponding bandwidths. After reconfiguring the network and providing a bandwidth guarantee by SDN controller, the VM migration plan is carried out at the corresponding time by OpenStack. By this way, it realizes the control and management of VM migrations.

To compute the migration sequence of VMs, we need network topology and traffic matrix of the data center. Besides, memory sizes and page dirty rates of VMs, and
residual physical resources such as CPU and memory are also needed.
Most of them can be obtained directly from the SDN controller or OpenStack,
but measurements of page dirty rate and traffic matrix
need special functions of the platform.
We next present the approach to measure them in details:

\noindent\textbf{Page Dirty Rate Measurement:}
We utilize
a mechanism called shadow page tables provided by Xen \cite{30} to track dirtying statistics on all pages \cite{1}.
All page-table entries (PTEs) are initially read-only mappings in the shadow tables.
Modifying a page of memory would result a page fault and then it is trapped by Xen.
If write access is permitted, appropriate bit in the VMs dirty bitmap is set to 1.
Then by counting the dirty pages in an appropriate period, we obtain the page dirty rate.

\begin{figure}[t]
\centering
\subfigure[Google's inter-datacenter WAN]{\includegraphics[width=.35\textwidth]{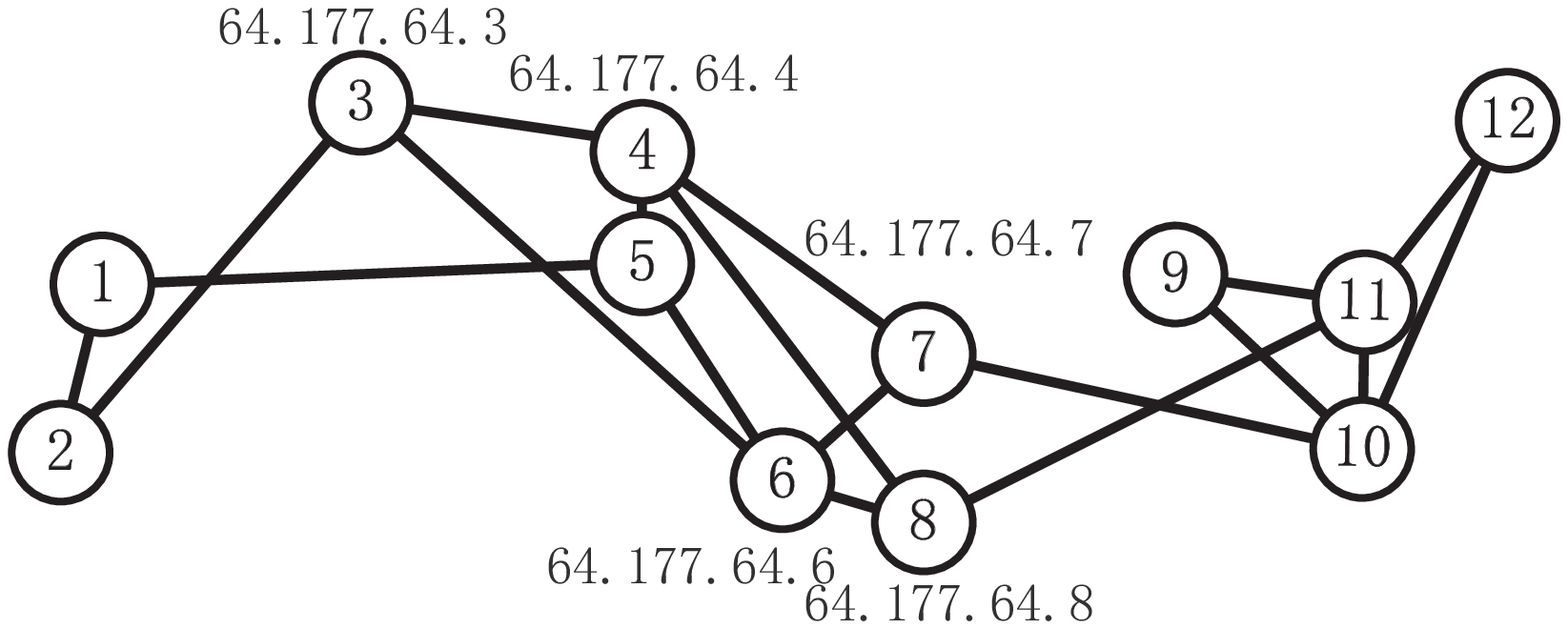}}
\subfigure[Modified Forwarding Table]{
\begin{minipage}{0.45\textwidth}
\centering
{\renewcommand\arraystretch{.3}
\begin{tabular*}{8cm}{c}
\tiny
\\
\end{tabular*}}
\footnotesize
\begin{tabular}{|c|c|c|c|}
\hline
{Prefix} & {Node} & {Next Hop} & {Traffic}  \\
\hline
195.112/16 & 64.177.64.8 & 64.177.64.6    &   $\alpha$    \\
\hline
195.027/16 & 64.177.64.3 & 64.177.64.4    &   $\beta$    \\
\hline
...&... &...& ...\\
\hline
\end{tabular}
\renewcommand\arraystretch{.3}
\begin{tabular*}{8cm}{c}
\tiny
\\
\end{tabular*}
\end{minipage}
}
\caption{Google's inter-datacenter WAN and the modified forwarding table for example.
}\label{fig:FT}
\end{figure}

 \noindent\textbf{Traffic Measurement:}
We assume SDN elements, switches and routers, can spilt traffic on multiple next hops correctly, and perform traffic measurements at the same time \cite{18,32}.
To aid traffic measurements,
an extra column in the forwarding table is used to record
the node in the network that can reach the destination IP address as in work \cite{18}.
 Take Fig.~\ref{fig:FT}(a), which is the topology of the inter-datacenter WAN of google, as an example,
 where all nodes are SDN forwarding elements.
 For instance, we assume node 8 (IP address 64.177.64.8) is the node that can reach the subset 195.112/16,
 and the shortest path from node 7 to node 8 goes through node 6 (IP address 64.177.64.6).
 Then, the forwarding table of node 7 is shown in Fig.~\ref{fig:FT}(b),
 where the first entry is corresponding to the longest matched prefix 195.112/16.
 When a packet with the longest matched prefix 195.112/16 is processed by node 7,
 $\alpha$ showed in the figure increases by the packet length.
 Thus, it tracks the number of bytes routed from node 7 to node 8 with the longest matched prefix 195.112/16.
 Using these data, the SDN controller easily obtains the traffic between arbitrary two nodes
 as well as residual capacity of each link.

\subsection{Problem Overview}\label{sec:PO}

In the system,
we assume there is no alternate network dedicated to VM migrations,
because of the cost of its deployment, especially in large-scale infrastructures.
Thus, only residual bandwidth can be used to migrate VMs.
Then, our goal is to determining the VMs' migration orders
and transmission rates that satisfy various constraints,
such as capacity constraints for memory and links,
to optimize the total migration time.

\begin{figure} [t]
\begin{center}
\includegraphics*[width=6cm]{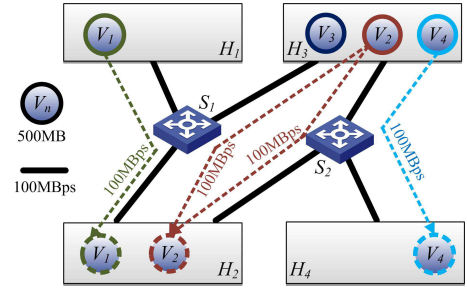}
\end{center}
\caption{An example of migration request and plan.} \label{fig:MRE}
\end{figure}

Now we give an example in Fig.~\ref{fig:MRE}.
In this network, there are 2 switches ($S_1$ and $S_2$) and
4 physical machines ($H_1$ to $H_4$) hosting 4 VMs ($V_1$ to $V_4$).
Assume the capacity of each link is 100MBps and memory size of each VM is 500MB.
We want to migrate $V_1$ from $H_1$ to $H_2$, $V_2$ from $H_2$ to $H_3$, and $V_4$ from $H_3$ to $H_4$.
The optimal plan of migration orders and transmission rates
is that first migrate $V_1$ and $V_4$ simultaneously,
respectively with paths $\{(H_1,S_1,H_2)\}$ and \{($H_3$,$S_2$,$H_4$)\}
and the corresponding maximum bandwidths of 100MBps.
Then migrate $V_2$ with paths $\{(H_2,S_1,H_3)$,\ $(H_2,S_2,H_3)$\}
and the corresponding maximum bandwidth of 200MBps.
It totally takes 7.5s to finish all the migrations.
Then, take random migration orders for example, $i.e.$,
first migrate $V_1$ and $V_2$ simultaneously,
respectively with paths $\{(H_1,S_1,H_2)\}$ and $\{(H_2,S_2,H_3)\}$
and the corresponding maximum bandwidths of 100MBps.
Then migrate $V_3$ with path $\{(H_3,S_2,H_4)\} $
and the corresponding maximum bandwidth of 100MBps.
It totally takes 10s to finish all the migrations.

In this example, $V_1$ and $V_4$ can be migrated in parallel,
while $V_2$ can be migrated with multipath.
However, $V_1$ and $V_2$, $V_4$ and $V_2$ share same links in their paths, respectively.
By determining a proper order, these migrations can be implemented
making full use of the network resources.
Thus, the total migration time is reduced by 25\% in the example,
illustrating the effect of the migration plan.

\subsection{Mathematical Model for Live Migration}\label{sec:Lmodel}

\begin{figure} [t]
\setlength{\abovecaptionskip}{0pt}
\setlength{\belowcaptionskip}{-10pt}
\begin{center}
\includegraphics*[width=8cm]{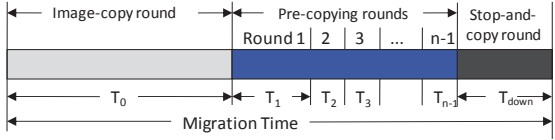}
\end{center}
\caption{Illustration of live migration performing pre-copy in iterative rounds.} \label{fig:LM}
\end{figure}

In this section, we present the mathematical model of live migration, which is presented in~\cite{5}.
We use $M$ to represent the memory size  of the virtual machine.
Let $R$ denote the page dirty rate during the migration and $L$ denote the bandwidth allocated for the migration.
Then, the process of the live migration is shown in Fig.~\ref{fig:LM}.
As we can observe, live migration copies memory in several rounds.
Assume it proceeds in $n$ rounds, and the data volume transmitted at each round is denoted by $V_i$ $(0\leq i\leq n)$.
At the first round, all memory pages are copied to the target host,
and we have $V_0=M$. Then in each round, pages that have been modified in the previous round are copied to the target host. The transmitted data can be calculated as $V_i=R\cdot T_{i-1},\ i>0$.
Thus, the elapsed time at each round can be calculated as
$T_i=V_i/L=R\cdot T_{i-1}/L=M\cdot R^i/L^{i+1}.$

Let $\lambda$ denote the ratio of $R$ to $L$, that is
$\lambda = R/L.$
Combining the above analysis, the total migration time can be represented as:
\begin{eqnarray}\label{equ:Tm}
\setlength{\abovedisplayskip}{3pt}
\setlength{\belowdisplayskip}{3pt}
T_{mig}=\sum_{i=0}^{n}T_i=\frac{M}{L}\cdot \frac{1-\lambda ^{n+1}}{1-\lambda}.
\end{eqnarray}

Let $V_{thd}$ denote the threshold value of the remaining dirty memory
that should be transferred at the last iteration.
We can calculate the total rounds of the iteration
by the inequality $V_n\leq V_{thd}$. Using the previous equations we obtain:
\begin{eqnarray}\label{equ:Rn}
\setlength{\abovedisplayskip}{3pt}
\setlength{\belowdisplayskip}{3pt}
n=\left \lceil log_\lambda \frac{V_{thd}}{M} \right \rceil.
\end{eqnarray}

In this model, the downtime caused in the migration can be represented as $T_{down}=T_d+T_r$,
where $T_d$ is the time spent on transferring the remaining dirty pages, and $T_r$ is the time spent on resuming the VM at the target host.
For simplicity, we assume the size of remaining dirty pages is equal to $V_{thd}$.


\subsection{Problem Formulation}

The network is represented by a graph $G=(V,E)$, where $V$ denotes the set of network nodes
and $E$ denotes the set of links.
Let $c(e)$ denote the residual capacity of the link $e\in E$.
Let a migration tuple $(s_k, d_k, m_k, r_k)$ denote that
a virtual machine should be migrated from the node $s_k$ to the node $d_k$
with the memory size $m_k$ and the page dirty rate $r_k$.
There are totally $K$ migration tuples in the system.
For the migration $k$, $l_k$ represents the bandwidth allocated for it.
Let $P_k$ denote the set of paths between $s_k$ and $d_k$.
The flow in path $p$ is represented by the variable $x(p)$.
Besides, as different migrations are started at different times,
we define binary variable $X_k$ to indicate
whether migration $k$ has been started at the current time.

We first discuss the optimization objective.
To obtain an expression of the total migration time is difficult in our model,
because we allow multiple VMs to be migrated simultaneously.
Thus, the total migration time cannot simply be represented
as the sum of the migration time of each VM like work \cite{12,13},
whose migration plans were designed under the model of one-by-one migration.
Moreover, even though we obtain the expression of the total migration time,
the optimization problem is still difficult and cannot be solved efficiently.
For example, work \cite{11} gives an expression of the total migration time
by adopting a discrete time model. However, they did not solve the problem directly,
instead, they proposed a heuristic algorithm independent with the formulation without any theoretical bound.
Thus, we try to obtain the objective function reflecting
the total migration time from other perspectives.

On the other hand,
since the downtime of live migration is required to be unnoticeable by users,
the number of the remaining dirty pages in the stop-and-copy round, $i.e.\ V_n$, need to be small enough.
According to the model provided in the last subsection, we have $V_n=M\cdot \lambda ^n$.
Thus, $\lambda ^n$ must be small enough.
For example, if migrating a VM, whose memory size is 10GB, with the transmission rate of 1GBps,
to reduce the downtime to 100ms, we must ensure $\lambda ^n\leq 0.01$.
Thus, by ignoring $\lambda ^n$ in the equation~(\ref{equ:Tm}), we have:
\begin{eqnarray}\label{equ:ATm}
\setlength{\abovedisplayskip}{3pt}
\setlength{\belowdisplayskip}{3pt}
T_{mig}\approx \frac{M}{L}\cdot \frac{1}{1-\lambda}=\frac{M}{L-R}.
\end{eqnarray}
We call the denominator as $net\ transmission\ rate$.
From an overall viewpoint, the sum of memory sizes of VMs is reduced
with the speed of $ \sum_{k=1}^{K}(l_k-X_k r_k)$,
which is the total net transmission rate in the network.
In turn, the integration of the net transmission rate
respect to time is the sum of memory sizes.
By maximizing the total net transmission rate,
we can reduce the total migration time efficiently.
Thus, it is reasonable for us to
convert the problem of reducing the migration time
to maximizing the net transmission rate,
which is expressed as $\sum_{k=1}^{K}(l_k-X_k r_k)$.

We now analyze constraints of the problem.
A VM is allowed to be migrated with multipath in our model.
Thus, we have a relationship between $l_k$ and $x(p)$:
$$
\setlength{\abovedisplayskip}{3pt}
\setlength{\belowdisplayskip}{3pt}
\sum_{p\in P_k} x(p) =l_k, \ \ \ k=1,...,K.
$$
Besides, the total flow along each link must not exceed its capacity. Thus, we have:
$$
\setlength{\abovedisplayskip}{3pt}
\setlength{\belowdisplayskip}{3pt}
\sum_{p\in P_e} x(p) \leq c(e), \ \ \forall e \in E.$$
For a migration that has not been started,
there is no bandwidth allocated for it.
Thus, we have constraints expressed as follow:
$$
\setlength{\abovedisplayskip}{3pt}
\setlength{\belowdisplayskip}{3pt}
 l_k \leq \beta \cdot X_k, \ \ \ k=1,...,K,$$
where $\beta$ is a constant large enough so that the maximum feasible bandwidth allocated for each migration cannot exceed it.
Then, the problem of maximizing the net transmission rate can be formulated as follows:
\begin{equation}\label{equ:prim}
\setlength{\abovedisplayskip}{3pt}
\setlength{\belowdisplayskip}{3pt}
\begin{array}{l}
\textbf{max}\ \sum_{k=1}^{K}(l_k-X_k r_k) \\
\textbf{s.t.}\ \
\begin{cases}
& \sum_{p\in P_k} x(p) =l_k, \ \ \ k=1,...,K \\
&\sum_{p\in P_e} x(p) \leq c(e), \ \ \forall e \in E\\
& l_k \leq \beta \cdot X_k, \ \ \ k=1,...,K \\
& X_k \in \{0,1\},\ \ \ k=1,...,K\\
& x(p)\geq 0, \ \ \ p\in P
\end{cases}
\end{array}
\end{equation}
which is a mixed integer programming (MIP) problem.

When some new migration requests come or old migrations are finished,
the input of the problem changes.
Thus, we recalculate the programming
under the new updated input. We notice that migrations that have been started cannot be stopped.
Otherwise, these migrations must be back to square one
because of the effect of the page dirty rate.
Thus, when computing this problem next time,
we add the following two constraints to it:
\begin{equation}\label{equ:Aprim}
\setlength{\abovedisplayskip}{3pt}
\setlength{\belowdisplayskip}{3pt}
\begin{array}{l}
\begin{cases}
X_k \geq X^0_k,\ \ \ k=1,...,K\\
l_k \geq l^0_k,\ \ \ k=1,...,K
\end{cases}
\end{array}
\end{equation}
where $X^0_k$ and $l^0_k$ are equal to the value of $X_k$ and $l_k$ in the last computing, respectively. It means a migration cannot be
stopped and its bandwidth does not decrease.

By solving the programming, we obtain the VMs
that should be migrated with their corresponding transmission rates,
maximizing the total net transmission rate under the current condition.
By dynamically determining the VMs to be migrated
in tune with changing traffic conditions and migration requests,
we keep the total net transmission rate maximized,
which is able to significantly reduce the total migration time.

\section{Approximation Algorithm} \label{sec:Approximation}

Solving the formulated MIP problem,
we obtain a well-designed sequence of the VMs to be migrated with their corresponding bandwidths.
However, the MIP problem is NP-hard,
and the time to find its solution is
intolerable on large scale networks.
For example, we implement the MIP problem using YALMIP --
a language for formulating generic optimization problems~\cite{17},
and utilize the GLPK to solve the formulation~\cite{21}.
Then, finding the solution of a network with 12 nodes and 95 VMs to be migrated
on a Quad-Core 3.2GHz machine takes at least an hour.
Therefore, we need an approximation algorithm with much lower time complexity.

\subsection{Approximation Scheme}
\subsubsection{Linear Approximation}
Let us reconsider the formulated MIP problem~(\ref{equ:prim}).
In this problem, only $X_k,\ k=1,...,K,$ are integer variables. Besides, the coefficient of $X_k$ in the objective function is $r_k$.
In practical data center, $r_k$ is usually much less than $l_k$, $i.e.$, the migration bandwidth of the VM.
Thus, we ignore the part of $\sum_{k=1}^{K}X_k r_k$ in the objective function, and remove variables $X_k,\ k=1,...,K$.
Then, we obtain a linear programming (LP) problem as follows:

\begin{equation}\label{equ:LP}
\setlength{\abovedisplayskip}{3pt}
\setlength{\belowdisplayskip}{3pt}
\begin{array}{l}
\textbf{max}\ \sum_{k=1}^{K}l_k \\
\textbf{s.t.}\ \
\begin{cases}
& \sum_{p\in P_k} x(p) =l_k, \ \ \ k=1,...,K \\
&\sum_{p\in P_e} x(p) \leq c(e), \ \ \forall e \in E\\
& x(p)\geq 0, \ \ \ p\in P
\end{cases}
\end{array}
\end{equation}
We select the optimal solution $l^*$ for~(\ref{equ:LP}) with most variables that are equal to zero
as our approximate solution.
Then we let $N^*$ denote the number of variables that are not zero in our approximate solution $l^*$,
and the corresponding binary decision variables $X_k$ are then set to be 1, while the other binary decision variables are set to be 0. Then the final approximate solution is denoted by $(l_k^*,X_k^*)$.

As for the primary problem with the additional constraints shown in~(\ref{equ:Aprim}), by a series of linear transformations, the problem is converted to a LP problem with the same form as~(\ref{equ:LP}) except for a constant in the objective function, which can be ignored.
Thus we obtain a linear approximation for the primary MIP problem.

\subsubsection{Fully Polynomial Time Approximation}
The exact solution of the LP problem~(\ref{equ:LP}) still cannot be found in polynomial time,
which means unacceptable computation time for large scale networks.
Thus, we further propose an algorithm to obtain the solution in polynomial time at the cost of accuracy.

Actually, ignoring the background of our problem
and removing the intermediate variable $l_k$,
we can express the LP problem~(\ref{equ:LP}) as:
\begin{equation}\label{equ:LPS}
\setlength{\abovedisplayskip}{3pt}
\setlength{\belowdisplayskip}{3pt}
\begin{array}{l}
\textbf{max}\ \sum_{p\in P}x(p) \\
\textbf{s.t.}\ \
\begin{cases}
&\sum_{p\in P_e} x(p) \leq c(e), \ \ \forall e \in E\\
& x(p)\geq 0, \ \ \ p\in P
\end{cases}
\end{array}
\end{equation}
This is a maximum multicommodity flow problem,
that is, finding a feasible solution for a multicommodity flow network that maximizes the total throughput.

\begin{algorithm}[bp!]
    \DontPrintSemicolon
  \caption{FPTA Algorithm.}
  \textbf{Input:} network $G(V,E)$, link capacities $c(e)$ for $\forall e\in E$, migration requests $(s_j,d_j)$  \\
  \textbf{Output:} Bandwidth $l_k$, binary decision variable $X_k$ for each migration $k$,
  and the amount of flow $x(p)$ in path $p\in P$.\\
  Initialize $u(e)=\delta \ \forall e\in E,\ x(p)=0\ \forall p\in P$\\
\For{$r=1$ \textup{to} $\left \lceil log_{1+\epsilon}\frac{1+\epsilon}{\delta}\right \rceil $}
{
    \For{$j=1$ \textup{to} $K$}
    {
        $p \leftarrow$ shortest path in $\mathcal{P}_j$\\
        \While{$u(p)<\min\{1,\delta (1+\epsilon)^r\}$}
        {
            $c\leftarrow \min_{e\in p}c(e)$\\
            $x(p)\leftarrow x(p)+c$\\
            $\forall e\in p, u(e)\leftarrow u(e)(1+\frac{\epsilon c}{c(e)})$\\
            $p \leftarrow$ shortest path in $\mathcal{P}_j$\\
        }
    }
}
\For{\textup{each} $p \in P$}
{
    $x(p)=x(p)/$log$_{1+\epsilon}\frac{1+\epsilon}{\delta}$\\
}
\For{$j=1$ \textup{to} $K$}
{
    $l_j=\sum_{p\in P_j}x(p)$\\
    $X_j=0$\\
    \If{$l_j\not=0$}
    {
        $X_j=1$\\
    }
}

$\mathbf{Return}\ (l_k,X_k)$ and $x(p)$
\end{algorithm}

Fleischer $et\ al.$\cite{9} proposed a Fully Polynomial-time Approximation Scheme (FPTAS) algorithm
independent of the number of commodities $K$ for the maximum multicommodity flow problem.
It can obtain a feasible solution whose objective function value is within $1+\epsilon$ factor of the optimal,
and the computational complexity is at most a polynomial function of the network size and $1/ \epsilon$.

Specifically, the FPTAS algorithm is a primal-dual algorithm.
We denote $u(e)$ as the dual variables of this problem.
For all $e\in E$, we call $u(e)$ as the length of link $e$.
Then, we define dist$(p)=\sum_{e\in p}u(e)$ as the length of path $p$.
This algorithm starts with initializing $u(e)$ to be $\delta$ for all $e\in E$ and $x(p)$ to be 0 for all $p\in P$.
$\delta$ is a function of the desired accuracy level $\epsilon$, which is set to be $(1+\epsilon)/((1+\epsilon)n)^{1/\epsilon }$ in the algorithm.
The algorithm proceeds in phases, each of which is composed of $K$ iterations.
In the $r_{th}$ phase, as long as there is some $p\in P_k$ for some $k$ with dist$(p)<$min$\{\delta(1+\epsilon)^r,1\}$,
we augment flow along $p$ with the capacity of the minimum capacity edge in the path.
The minimum capacity is denoted by $c$. Then, for each edge $e$ on $p$,
we update $u(e)$ by $u(e)=u(e)(1+\frac{\epsilon c}{c(e)})$.
At the end of the $r_{th}$ phase, we ensure every $(s_j, d_j)$ pair is at least $\delta(1+\epsilon)^r$ or $1$ apart.
When the lengths of all paths belonging to $P_k$ for all $k$ are between $1$ and $1+\epsilon$, we stop.
Thus, the number of phases is at most $\left \lceil log_{1+\epsilon}\frac{1+\epsilon}{\delta}\right \rceil$.
Then, according to theorem in \cite{9}, the flow obtained by scaling the final flow obtained in previous phases by log$_{1+\epsilon}\frac{1+\epsilon}{\delta}$ is feasible.
We modified the FPTAS algorithm by adding some post-processes to obtain
the feasible $(l_k,X_k)$ and $x(p)$ to the primal MIP problem,
and the modified algorithm is given in more detail in Algorithm 1.
The computational complexity of the post-processes is only a linear function of the number of the VMs to be migrated.
In addition, the computational complexity of the FPTAS algorithm is at most a polynomial function of the network size and $1/ \epsilon$ \cite{9}.
Thus, the computational complexity of our approximation algorithm is also polynomial.
and we obtain a fully polynomial time approximation (termed as FPTA)
to the primal MIP problem.

\subsection{Bound  Analysis}\label{sec:BA}
To demonstrate the effectiveness of our proposed algorithm,
we now analyze the bound of it.
We first analyze the bound of the linear approximation compared with the primary MIP problem (\ref{equ:prim}),
then analyze the bound of the FPTA algorithm compared with the linear approximation (\ref{equ:LP}).
With these two bounds, we finally obtain the bound of the FPTA algorithm showing in Algorithm 1 compared with the primary MIP problem (\ref{equ:prim}).

\subsubsection{Bound of the Linear Approximation}\label{sec:BLAS}
We discuss the bound of the linear approximation
compared with the primary MIP problem in normal data center network scenarios.
Common topologies of data center networks, such as fat tree, usually provide full bisection bandwidth,
which enables all hosts communicating with each other with full bandwidth at the same time.
Thus, we can ignore the routing details, and only guarantee the traffic at each host not exceeds its maximum bandwidth.
Then, the LP problem (\ref{equ:LP}) becomes:
\begin{equation}\label{equ:LPR}
\setlength{\abovedisplayskip}{3pt}
\setlength{\belowdisplayskip}{3pt}
\begin{array}{l}
\textbf{max}\ \sum_{k=1}^{K}l_k \\
\textbf{s.t.}\ \
\begin{cases}
& \sum_{s_k=i} l_k \leq C^s_i, \ \ \ i=1,...,H \\
& \sum_{d_k=i} l_k \leq C^d_i, \ \ \ i=1,...,H \\
& l_k\geq 0, \ \ \ k=1,...,K
\end{cases}
\end{array}
\end{equation}
where $C^s_i$ is the maximum amount of traffic that can be received at host $i$,
while $C^d_i$ is the maximum amount of traffic that can be sent at host $i$.
Besides, there are $H$ hosts in the data center.
Then, we let $L_0$ be the minimum of $C^s_i$ and $C^d_i$.
That is, min$\{C^s_1,...,C^s_H\}\geq L_0$ and min$\{C^d_1,...,C^d_H\}\geq L_0$.
Similarly, we let $R_0$ be the maximum of $r_k$. That is, max$\{r_1,...,r_K\}\leq R_0$.

We now provide some supplement knowledge about linear programming.
For a linear programming with standard form, which can be represented as:
\begin{equation}\label{equ:LPSF}
\setlength{\abovedisplayskip}{3pt}
\setlength{\belowdisplayskip}{3pt}
\begin{array}{l}
\textbf{max}\ b^T x \\
\textbf{s.t.}\ \
\begin{cases}
& \!\!\!\!\!\!Ax=c \\
& \!\!\!\!\!\!x\geq 0
\end{cases}
\end{array}
\end{equation}
where $x,b \in R^n$, $c\in R^m$, $A \in R^{m\times n}$ has full rank $m$, we have the following
definitions and lemmas.

\textbf{Definition 1 (Basic Solution)}
Given the set of $m$ simultaneous linear equations in $n$ unknowns of $Ax=c$ in (\ref{equ:LPSF}),
let $B$ be any nonsingular $m\times m$ submatrix made up of columns of $A$.
Then, if all $n-m$ components of $x$ not associated with columns of $B$ are set
equal to zero, the solution to the resulting set of equations is said to be a basic
solution to $Ax=c$ with respect to the basis $B$. The components of $x$ associated
with columns of $B$ are called basic variables, that is, $Bx_B=c$~\cite{23}.


\textbf{Definition 2 (Basic Feasible Solution)}
A vector $x$ satisfying (\ref{equ:LPSF}) is said to be feasible for these constraints.
A feasible solution to the constraints (\ref{equ:LPSF}) that is also basic is said to
be a basic feasible solution~\cite{23}.


\textbf{Lemma 1 (Fundamental Theorem of LP)}  Given a linear program in
standard form (\ref{equ:LPSF}) where $A$ is an $m\times n$ matrix of rank $m$.
If there is a feasible solution, there is a basic feasible solution.
If there is an optimal feasible solution, there is an optimal basic feasible
solution~\cite{23}.

These definitions and the lemma with its proof can be found
in the textbook of linear programming~\cite{23}.
With these preparations, we have the following lemma:

\textbf{Lemma 2} There exists an optimal solution for (\ref{equ:LPR}),
such that there are at least $N^*$ equalities that hold in inequality constraints of (\ref{equ:LPR}).

\textbf{Proof}: Problem (\ref{equ:LPR}) can be represented in standard form as:
\begin{equation}\label{equ:LPMFSF}
\setlength{\abovedisplayskip}{3pt}
\setlength{\belowdisplayskip}{3pt}
\begin{array}{l}
\textbf{max}\ b^T l \\
\textbf{s.t.}\ \
\begin{cases}
& \!\!\!\!\!\!\left[A\ I\right] \left[\begin{array}{cc}l \\ s \end{array}\right] =c \\
& \!\!\!\!\!\!l,s\geq 0
\end{cases}
\end{array}
\end{equation}
where $c\in R^{2H}$, $l = (l_1,l_2,...,l_K)^T,\ s = c-Al,\ b = (1,1,...,1)^T\in R^K$, $I \in R^{K\times K}$ is the identity matrix of the order $K$, $A \in R^{2H\times K}$ is composed of 0 and 1, and each column of $A$ has and only has two elements of 1.
Besides, $\left[A\ I\right] \in R^{2H\times K+2H}$ has full rank $2H$.

\begin{figure*}[t]
\centering
\subfigure[]{\includegraphics[width=.31\textwidth]{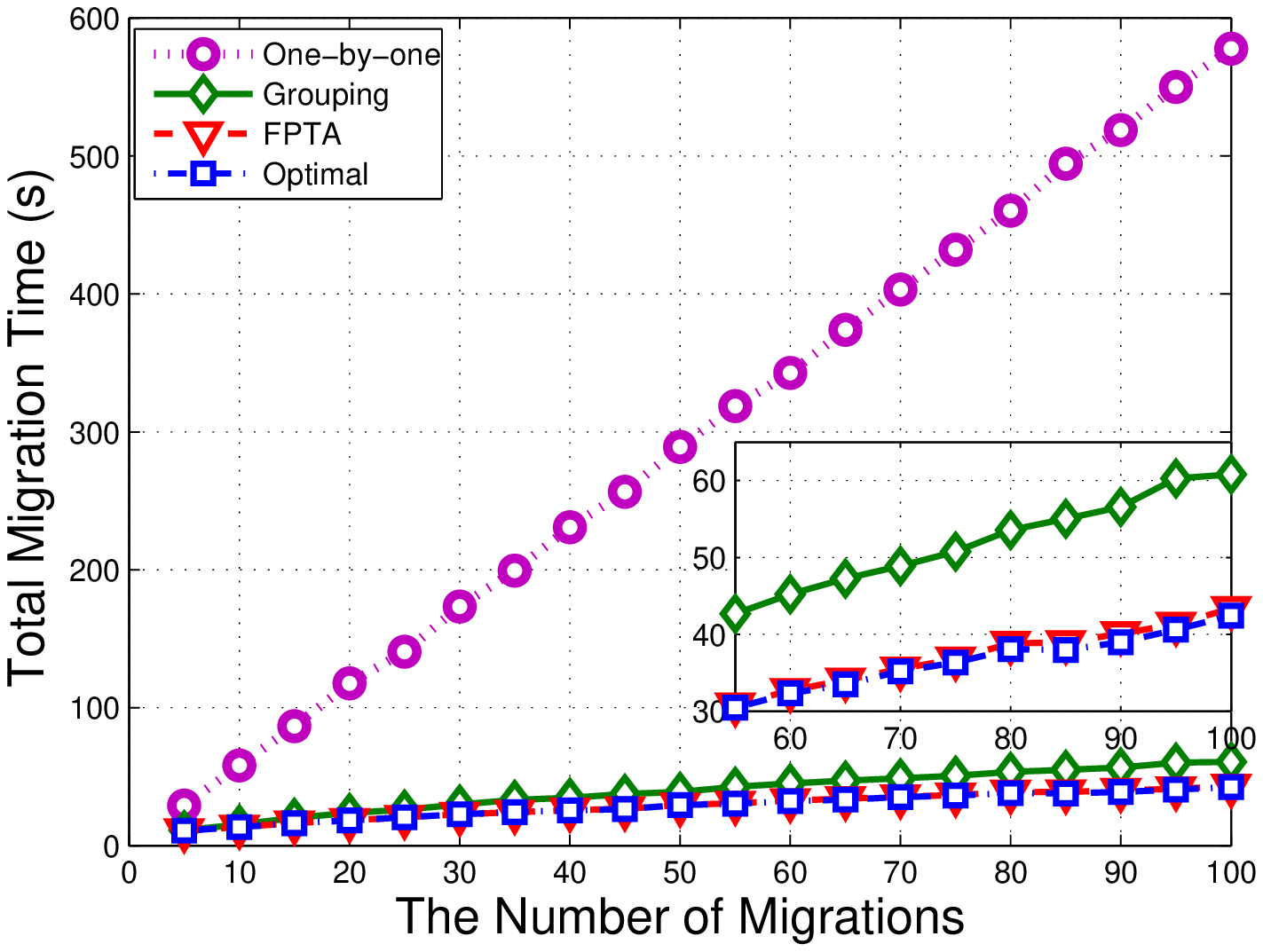}}\label{TMTvsDP1}
\hspace{0.1in}\subfigure[]{\includegraphics[width=.31\textwidth]{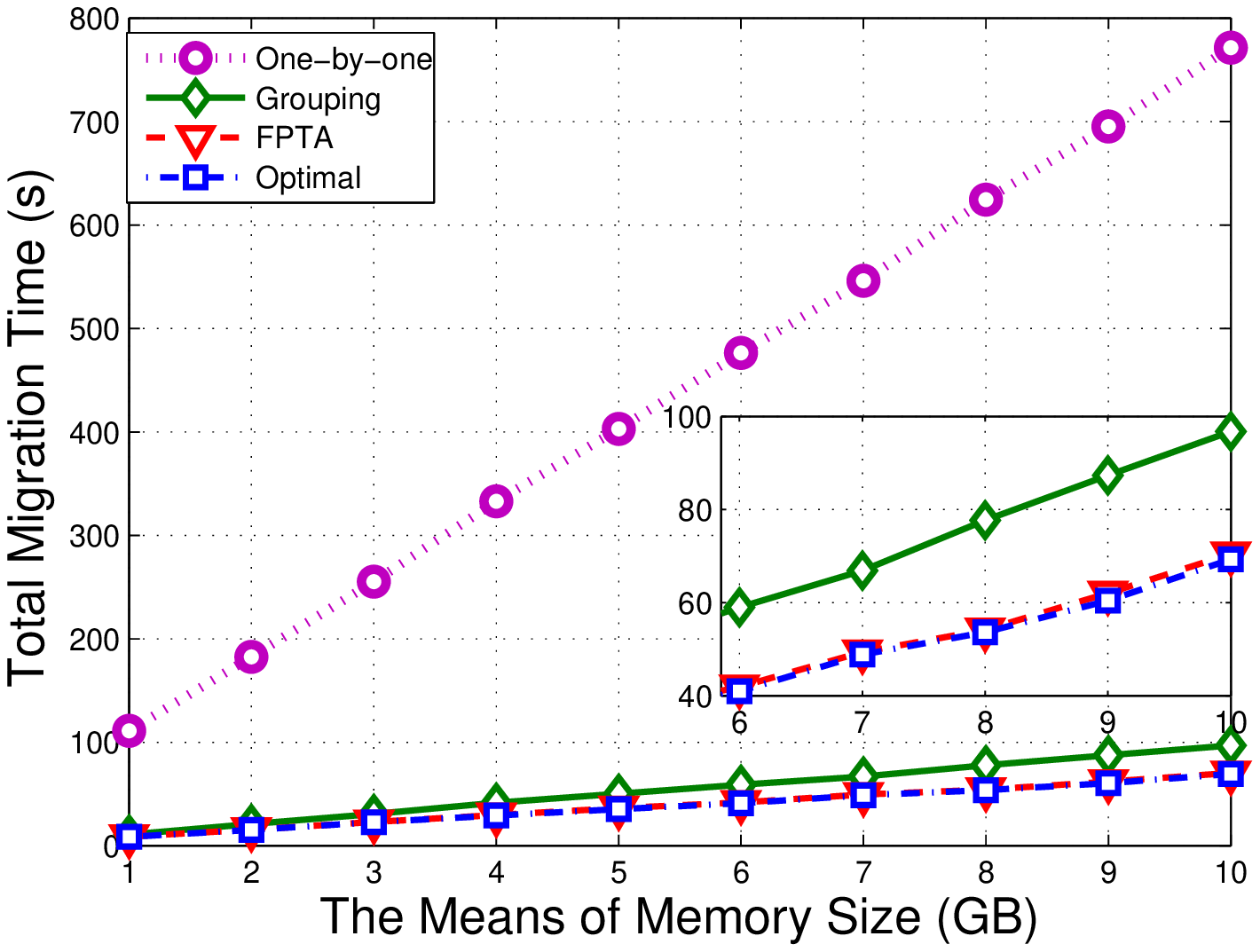}}
\hspace{0.1in}\subfigure[]{\includegraphics[width=.31\textwidth]{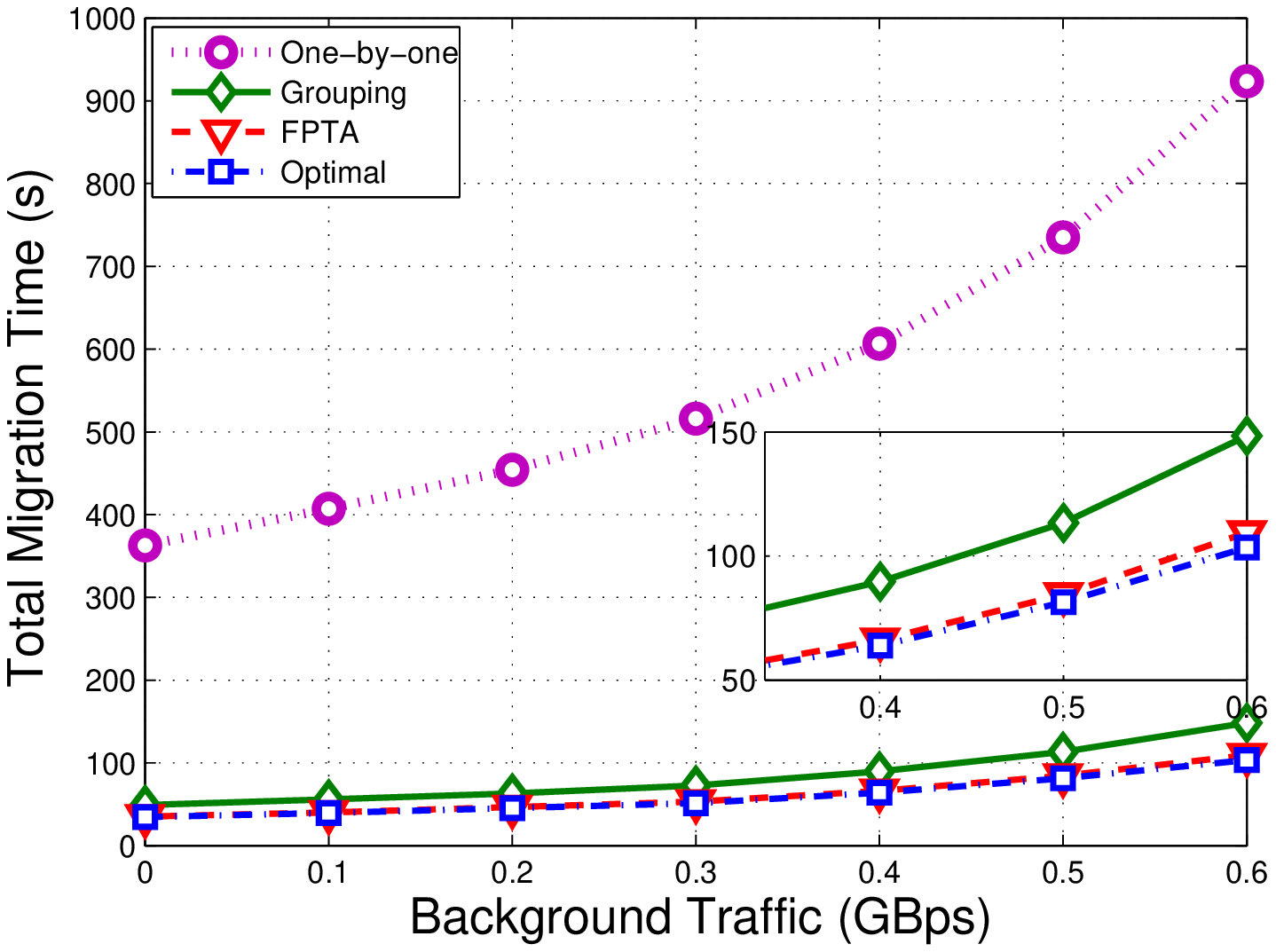}}
\caption{Total migration time vs different parameters in one datacenter
 under the topology of PRV1.} \label{TMTvsDP}
\end{figure*}

By Lemma 1, if the LP problem (\ref{equ:LPMFSF}) has an optimal feasible solution,
we can find an optimal basic feasible solution $(\hat{l},\hat{s})$ for (\ref{equ:LPMFSF}).
By the definition of basic solution, the number of nonzero variables in $(\hat{l},\hat{s})$ is less than $2H$.
Meanwhile, By the definition of $N^*$,
the number of nonzero variables in $\hat{l}$,
which is represented by $\hat{N}$, is greater than $N^*$.
Thus the number of nonzero variables in $\hat{s}$ is less than $2H-N^*$.
Then there are at least $N^*$ variables that are equal to zero in $\hat{s}$.
Meanwhile, $\hat{s}_j=0,\ j\in \{1,...,2H\}$ means the equality holds in the inequality constraint
corresponding $j_{th}$ row in $A$.
Therefore, we have at least $N^*$ equalities that hold in inequality constraints of (\ref{equ:LPR}).$\blacksquare$



\textbf{Theorem 1} Assume $R_0=\eta L_0$. Let $U$ be the optimal value of the primal MIP problem (\ref{equ:prim}), and $V$ be the optimal value of the LP problem (\ref{equ:LPR}).
Then we have $V-N^*R_0\geq (1-\sigma)U$, where $\sigma=\frac{2\eta}{1-2\eta}$.

\textbf{Proof}:
We first prove $V\geq \frac{1}{2}N^*L_0$.
By lemma 2, we know that there exists an optimal solution of (\ref{equ:LPR})
such that there are at least $N^*$ equalities that hold in inequality constraints of (\ref{equ:LPR}).
We select the corresponding rows $a_1,...a_{N^*}$ of $A$ and corresponding elements $c_1,...c_{N^*}$ of $c$.
Then we have $a^T_i \hat{l}=c_i,\ i=1,...,N^*$.
Because each column of $A$ has and only has two elements of 1,
elements of $\sum_{i=1}^{N^*}a_i$ are at most 2.
Thus, we have $V=\sum_{k=1}^{K}\hat{l}_k\geq \frac{1}{2}\sum_{i=1}^{N^*}a^T_i \hat{l}
=\frac{1}{2}\sum_{i=1}^{N^*}c_i\geq \frac{1}{2}N^*L_0$.

By definition of $U$ and $V$, we have $U\leq V$ and $V-N^*R_0\leq U$.
Then we have $\left|U-(V-N^*R_0)\right|=U-V+N^*R_0\leq N^*R_0$.
Besides, by the last paragraph, we have $U\geq V-N^*R_0\geq \frac{1}{2}N^*L_0-N^*R_0$.
Thus, we have $\frac{\left|U-(V-N^*R_0)\right|}{U}=\frac{U-(V-N^*R_0)}{U}
\leq \frac{N^*R_0}{\frac{1}{2}N^*L_0-N^*R_0}
=\frac{2R_0}{L_0-2R_0}=\frac{2\eta}{1-2\eta}=\sigma$, $i.e.$, $V-N^*R_0\geq (1-\sigma)U$.
$\blacksquare$

By the definitions of $N^*$ and $R_0$, we have that the net transmission rate corresponding to the selected solution of (\ref{equ:LPR}) is at least $V-N^*R_0$.
Thus, we obtain the bound of the linear approximation
compared with the primary MIP problem.

\subsubsection{Bound of the FPTA Algorithm}
We next analyze the bound of the FPTA algorithm.
According to theorem in \cite{9}, we have the following lemma:


\textbf{Lemma 3} 
If $p$ is selected in each iteration to be the shortest
$(s_i,d_i)$ path among all commodities, then for a final flow value $W=\sum_{p\in P}x(p)$
obtained from the FPTAS algorithm,
we have $W\geq (1-2\epsilon)V$, where $\epsilon$ is the desired accuracy level.

Because the value of $x(p)$ is unchanged in our post-processes of Algorithm 1,
$W$ is also the final flow value of our proposed FPTA algorithm.
Note that it is not the bound of the FPTA algorithm compared with the LP problem (\ref{equ:LP}), because our objective function is the net transmission rate,
while $W$ is only the transmission rate of the solution of the FPTA algorithm.
Besides, $V$ is not the maximum net transmission rate as well.
The bound of the FPTA algorithm is given in the following theorem:

\textbf{Theorem 2} Let $F$ be the net transmission rate corresponding to the solution of Algorithm 1. In the data center networks providing full bisection bandwidth,
we have $F\geq (1-2\epsilon-\sigma)U$, where $U$ is the optimal value
of the primal MIP problem (\ref{equ:prim}).

\textbf{Proof}:
By the definitions of $N^*$ and $R_0$, we have that the net transmission rate corresponding to the solution of the FPTA algorithm is at least $W-N^*R_0$, $i.e.$, $F\geq W-N^*R_0$.
Thus we have $F\geq (1-2\epsilon)V-N^*R_0=(1-2\epsilon)(V-N^*R_0)-2\epsilon N^*R_0$.
Meanwhile, by $U\geq \frac{1}{2}N^*L_0-N^*R_0\geq \frac{1}{2\eta}N^*R_0-N^*R_0$,
we have $N^*R_0\leq \frac{2\eta}{1-2\eta}U=\sigma U$.

By Theorem 1,
we have $F\geq (1-2\epsilon)(1-\sigma)U-2\epsilon \sigma U=(1-2\epsilon-\sigma)U$.
Thus we obtain the bound of the FPTA algorithm compared with the primal MIP problem.$\blacksquare$

\section{Performance Evaluation}\label{sec:Evaluation}
\subsection{Simulation System Set Up}

With the increasing trend of owning multiple datacenter sites by a single company,
migrating VMs across datacenters becomes a common scenario.
Thus, to evaluate the performance of our proposed migration plan inside one datacenter
and across datacenters, we select the following two topologies
to implement our experiments:
(1) The topology of a private enterprise data center located in Midwestern United States (PRV1 in
\cite{7}).
(2) B4, Google's inter-datacenter WAN with 12 data centers
interconnected with 19 links~\cite{6} (showing in Fig.~\ref{fig:FT}(a)).
In B4, each node represents a data center.
Besides, the network provides massive bandwidth.
However, to evaluate the performance of our proposed algorithm under relatively hard conditions,
we assume the capacity of each link is only 1GBps.
On the other hand,
the capacities of links in RPV1 are set ranging from 1GB to 10GB according to~\cite{7}.
The page dirty rate is set to 100MBps.
Besides, $V_{thd}$ and $T_r$ are set to 100MB and 20ms, respectively.
The memory sizes of VMs are also set ranging from 1GB to 10GB unless stated otherwise.

In our experiments, we evaluate the performance of our proposed FPTA algorithm
compared with the optimal solution of the MIP problem (referred to as optimal
algorithm) and two state-of-the-art algorithms.
In the two state-of-the-art algorithms,
one is the algorithm based on one-by-one migration scheme (referred to as one-by-one algorithm),
which is proposed in \cite{12,13}.
The other is the algorithm that migrates VMs by groups (referred to as grouping algorithm), just as the algorithm proposed in \cite{11}.
In this algorithm,
VMs that can be migrated in parallel are divided into the same group,
while VMs that share the same resources, such as the same link in their paths,
are divided into different groups.
Then VMs are migrated by groups according to their costs \cite{11}.
We further set the function of the cost as the weighted value of the total migration time and the number of VMs in each group.

\subsection{Results and Analysis}
\subsubsection{Migration Time}

\begin{figure}
\centering
\subfigure[]{\includegraphics[width=.23\textwidth]{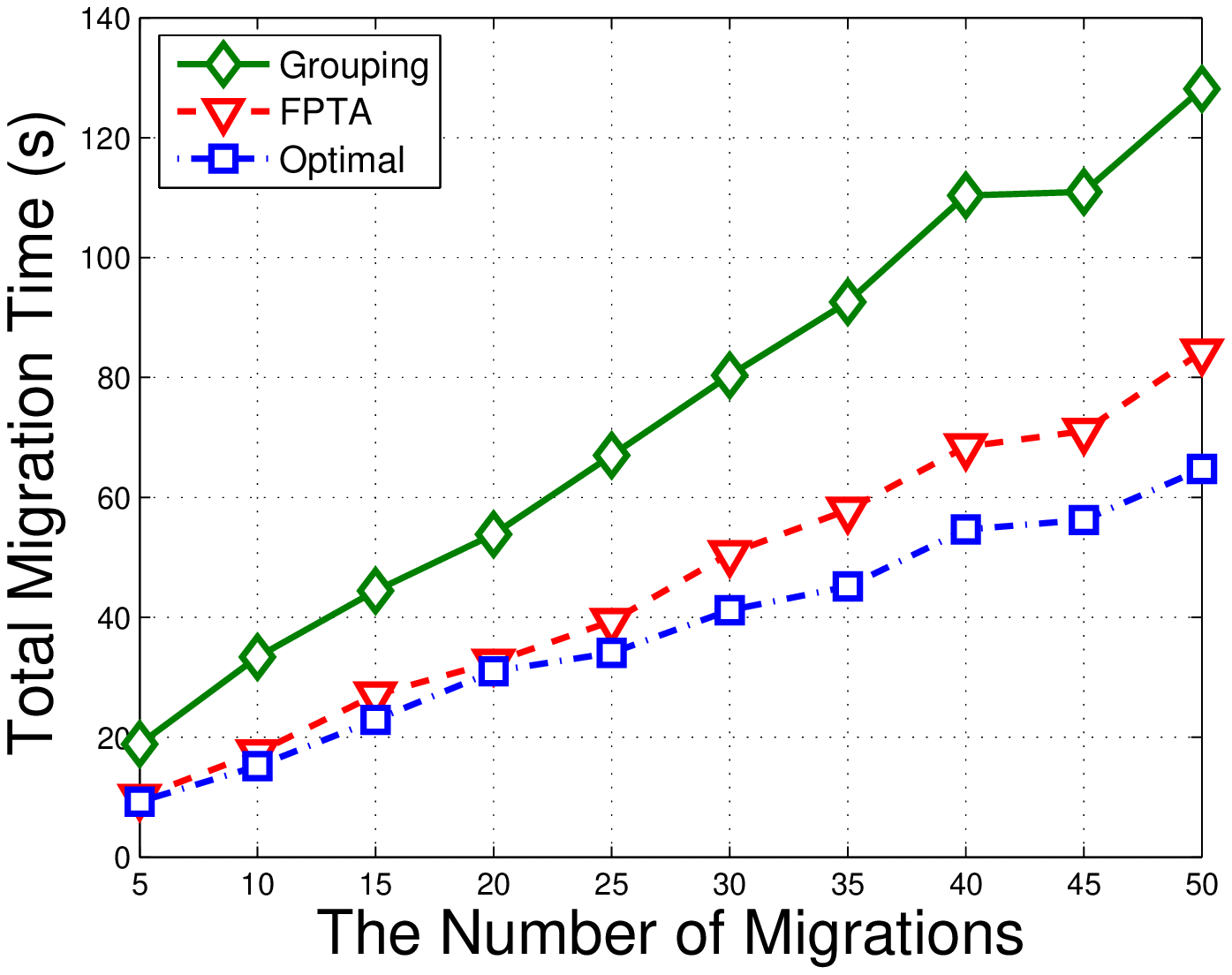}}
\hspace{0.05in}\subfigure[]{\includegraphics[width=.23\textwidth]{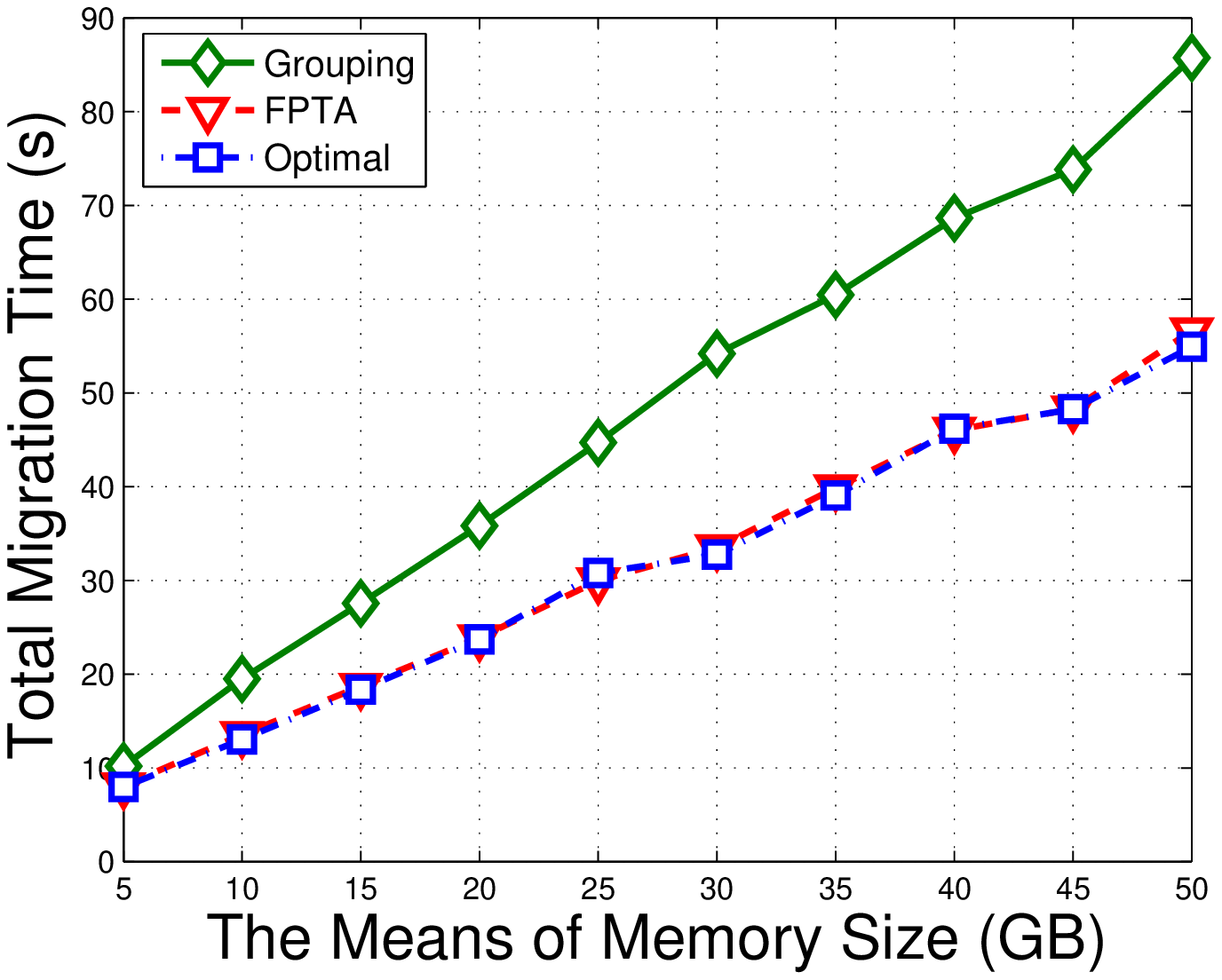}}
\hspace{0.05in}\subfigure[]{\includegraphics[width=.23\textwidth]{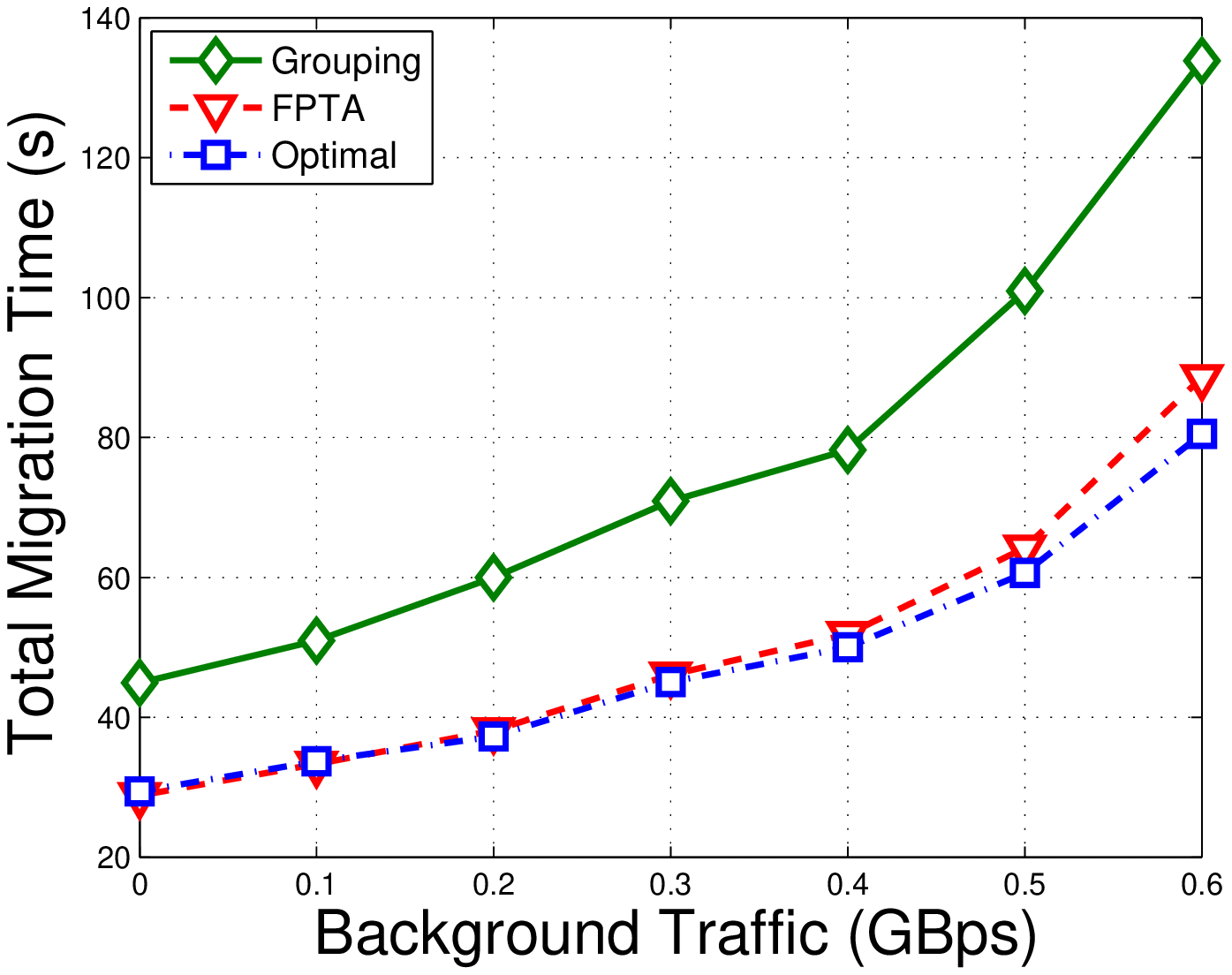}}
\hspace{0.05in}\subfigure[]{\includegraphics[width=.23\textwidth]{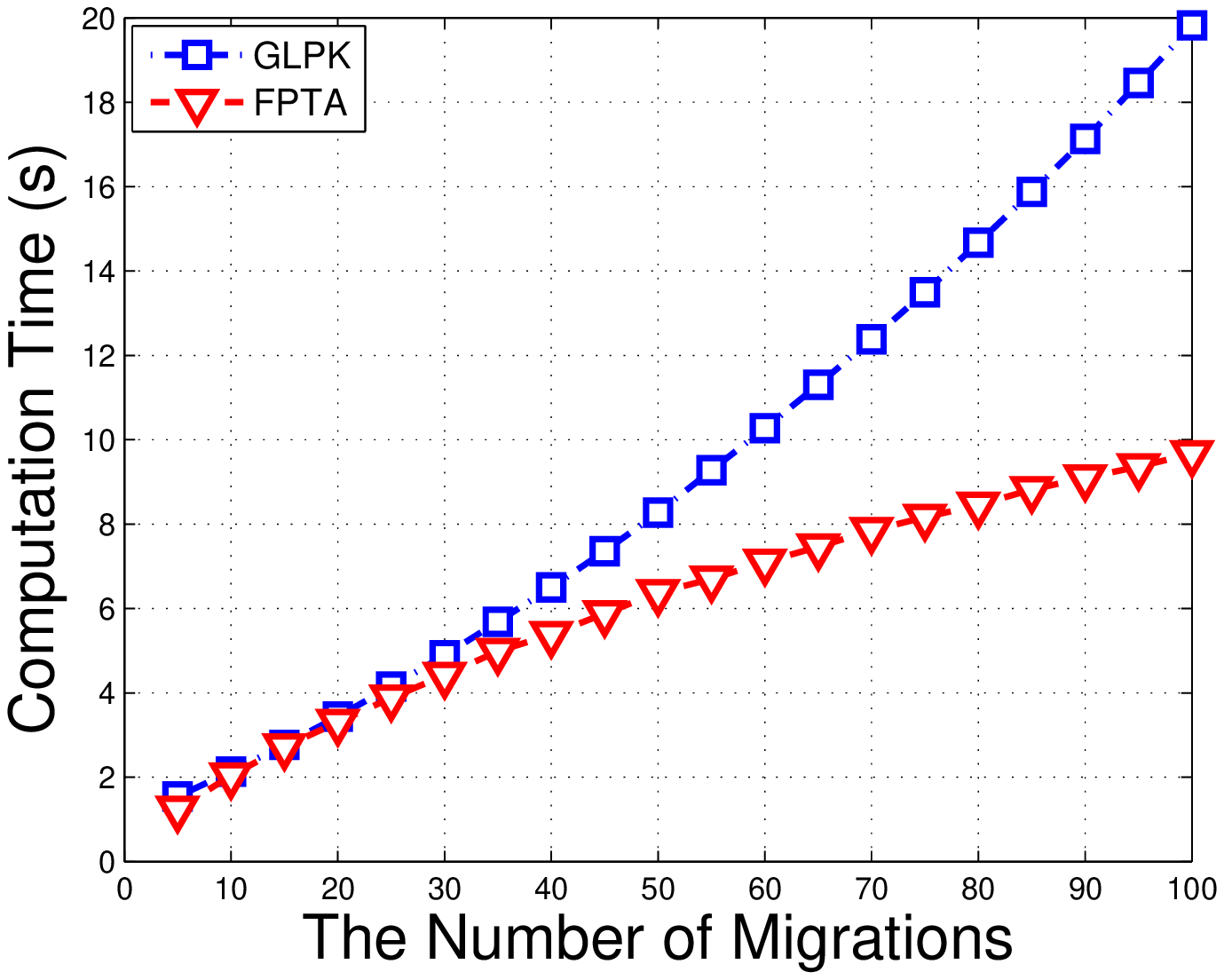}}
\caption{Total migration time or computation time vs different parameters in inter-datacenter network under the topology of B4.} \label{TMTvsDP2}
\end{figure}

In our first group of experiments,
we compare the total migration time of our proposed FPTA algorithm with
that of other algorithms introduced above, with the variation of
different parameters, $i.e.$, the number of VMs to be migrated, the amount of background traffic, the average memory size of VMs,
in PRV1 and B4, respectively.
The results are shown in Fig.~\ref{TMTvsDP} and Fig.~\ref{TMTvsDP2}.

As we can observe from the Fig.~\ref{TMTvsDP}, 
the performance of the one-by-one algorithm is much worse than that of the
other three algorithms: when there are 100 VMs to be migrated, the total migration time it takes is about 10 times more than that of the other three algorithms, illustrating its inefficiency
in reducing the total migration time.
Since the performance gap between one-by-one algorithm and the other algorithms is huge,
we do not show its performance in Fig.~\ref{TMTvsDP2}.
Besides, from Fig.~\ref{TMTvsDP2}(d) we can observe that
the computation time of our proposed FPTA algorithm is at most a polynomial function of the number of the migrations, much less than that of using GLPK to solve the LP problem (\ref{equ:LP}).

As for the performance of the other three algorithms,
their total migration time vs different parameters in data center networks and inter-datacenter WAN
has a similar trend: the total migration time of FPTA algorithm is very close to that of the optimal algorithm,
and much less than that of the grouping algorithm. Take Fig.~\ref{TMTvsDP}(a) and Fig.~\ref{TMTvsDP2}(a) for example.
In PRV1 (showing in Fig.~\ref{TMTvsDP}(a)), total migration time of the FPTA algorithm
and the optimal algorithm almost cannot be distinguished,
while in B4 (showing in Fig.~\ref{TMTvsDP2}(a)) the gap is less than 15\% relative to the optimal algorithm.
Meanwhile, FPTA algorithm performs much better than the grouping algorithm:
its migration time is reduced by 40\% and 50\% in comparison with the grouping algorithm
in PRV1 and B4, respectively.
Thus, the solution of our proposed FPTA algorithm approaches to the optimal solution
and outperforms the state-of-the-art solutions.

\subsubsection{Net Transmission Rate}

\begin{figure} [t]
\begin{center}
\includegraphics*[width=8.5cm]{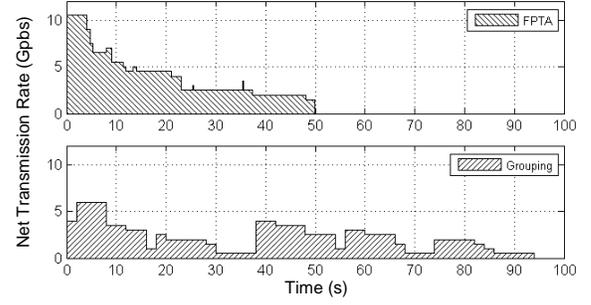}
\end{center}
\caption{The net transmission rates vs time of
FPTA algorithm and grouping algorithm in inter-datacenter network under the topology of B4
with 40 VMs to be migrated.} \label{fig:NTRvsT}
\end{figure}

To illustrate the effectiveness of maximizing the net transmission rate,
we implement the second group of experiments in the scenario
where there are 40 VMs to be migrated in B4.
Net transmission rates of the FPTA algorithm and the grouping algorithm are evaluated,
as functions of time. The result is shown in Fig.~\ref{fig:NTRvsT}.

According to previous theoretic analysis,
we know that the sum of memory sizes of VMs to be migrated
is approximately equal to the integration of the net transmission rate with respect to time.
In the experiments, the sum of memory sizes of the 40 VMs to be migrated are 203GB.
Meanwhile, in Fig.~\ref{fig:NTRvsT}, the shadow areas of the FPTA and grouping algorithm,
which can represent the integrations of the net transmission rates with respect to time,
are 203.0GB and 212.0GB, respectively. The relative errors are less than 5\%.
It proves the correctness of our theoretic analysis.
Besides, from the figure we observe that the net transmission rate with the FPTA algorithm
remains a relatively high level in the process of migrations,
about 2 times higher than that of the grouping algorithm on average.
Thus, the integration of the net transmission rate can reaches $\sum_{k=1}^{40}m_k$ with less time.
Specifically, in this group of experiments, the total migration time of FPTA algorithm is reduced by up to 50\% compared with grouping algorithm.
Thus, our FPTA algorithm significantly reduces the total migration time by maximizing the net transmission rate.

\subsubsection{Application Performance}
\begin{figure} [t]
\begin{center}
\includegraphics*[width=8.5cm]{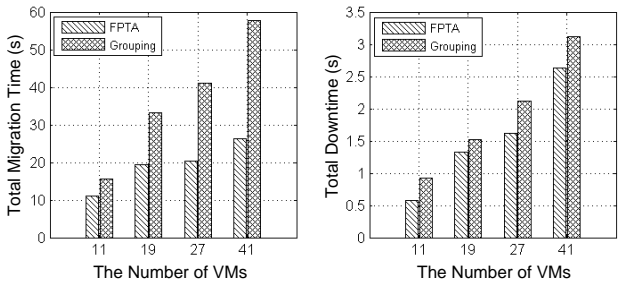}
\end{center}
\caption{Total migration time and downtime for optimizing delay
 in inter-datacenter network under the topology of B4.} \label{fig:RS}
\end{figure}

The scenarios of this group of experiments are to optimize the average delay of services in B4.
Assume there are some VMs located randomly in the data centers in B4 at the beginning,
and they are providing services to the same user,
who is located closely to the node 8 (data center 8).
Thus we need to migrate these VMs to data centers as close to the node 8 as possible.
However, memory that each data center provides is not unlimited,
which is set to be 50GB in our experiments.
Besides, there are 11, 19, 27, 41 VMs in the network, respectively.
We find the final migration sets by minimizing the average delay.
Then we use the FPTA and grouping algorithm to implement these migrations.
The results are shown in Fig.~\ref{fig:RS}.

Fig.~\ref{fig:RS}(a) and (b) show the total migration time and downtime, respectively.
As we observe,
FPTA algorithm reduces the total migration time and downtime by 43.7\% and 22.6\% on average
compared with those of the grouping algorithm, respectively.
Thus, our proposed FPTA algorithm outperforms the grouping algorithm uniformly,
which provides better services for the user.

\section{Related Work} \label{sec:RelatedWorks}
Works related to our paper can be divided by two topics:
live migration and migration planning.

Since Clark proposed live migration \cite{1},
there have been plenty of works that have been done in this field.
Ramakrishnan $et\ al.$ \cite{36} advocated a cooperative, context-aware approach to data center migration across WANs to deal with outages in a non-disruptive manner.
Wood $et\ al.$ \cite{37} presented a mechanism
that provides seamless and secure cloud connectivity
as well as supports live WAN migration of VMs.
On the other hand, VM migration in SDNs has made some progress.
Mann $et\ al.$ \cite{33} presented CrossRoads -- a network fabric that provides
layer agnostic and seamless live and offline VM mobility across
multiple data centers.
Boughzala $et\ al.$ \cite{10} proposed a network infrastructure based on OpenFlow
that solves the problem of inter-domain VM migration.
Meanwhile, Keller $et\ al.$ \cite{16} proposed LIME, a general and efficient solution
for joint migration of VMs and the network.
These works indicate that SDN has big advantages in implementing VM migration.
In contrast, we focus on developing a VM migration plan to reduce
the total migration time in Software Defined Network (SDN)
scenarios.

Meanwhile, there have been some works about VM migration planning.
However, most of them were designed under the model of one-by-one migration\cite{12,13}
or their main focuses were not to optimize the total migration time \cite{13,14}.
Ghorbani $et\ al.$ \cite{13} proposed a heuristic algorithm of determining the ordering of VM migrations and corresponding OpenFlow instructions.
However, they concentrated on bandwidth guarantees, freedom of loops,
and their algorithm is based on the model of one-by-one migration.
Al--Haj $et\ al.$ \cite{14} also focused on finding a sequence of migration steps.
Their main goal was to satisfy security, dependency, and performance requirements.

\section{Conclusion} \label{sec:Conclusion}
In this work,
we focus on reducing the total migration time
by determining the migration orders and transmission rates of VMs.
Since solving this problem directly is difficult,
we convert the problem to another problem, $i.e.$, maximizing the net transmission rate in the network.
We formulate this problem as a mixed integer programming problem, which is NP-hard.
Then we propose a fully polynomial time approximation (FPTA) algorithm to solve the problem.
Results show that the proposed algorithm
approaches to the optimal solution
with less than 10\% variation and much less computation time.
Meanwhile, it reduces the total migration time and the service
downtime by up to 40\% and 20\% compared with the state-of-the-art
algorithms, respectively.

\end{document}